\documentclass[nofootinbib,pra,eprint,twocolumn,showpacs,superscriptaddress,notitlepage,amsmath,floatfix]{revtex4-2} 

\usepackage[colorlinks=true,citecolor=blue,linkcolor=blue,urlcolor=blue]{hyperref}
\usepackage{amssymb}

\usepackage{dsfont}
\usepackage{txfonts}
\usepackage{framed}
\usepackage{braket}
\usepackage{svg}
\definecolor{shadecolor}{gray}{0.95}
\usepackage{makecell}
\usepackage{multirow}
\usepackage{bbold}
\usepackage[caption=false]{subfig}

\definecolor{newblue}{RGB}{112,178,255}
\definecolor{neworange}{RGB}{255,204,112}
\definecolor{blue2}{RGB}{120,0,255}
\definecolor{red2}{RGB}{255,0,120}
\definecolor{green2}{RGB}{0,130,130}

\newcommand{\half}{\frac{1}{2}}
\definecolor{darkred}{RGB}{245,186,183}
\definecolor{lightred}{RGB}{249,217,215}

\def\ket#1{\mathinner{|{#1}\rangle}}

\begin{document}

\begin{abstract}
Calculating the equilibrium properties of condensed matter systems is one of the promising applications of near-term quantum computing. Recently, hybrid quantum-classical time-series algorithms have been proposed to efficiently extract these properties from a measurement of the Loschmidt amplitude $\langle \psi| e^{-i \hat H t}|\psi \rangle$ from initial states $|\psi\rangle$ and a time evolution under the Hamiltonian $\hat H$ up to short times $t$. In this work, we study the operation of this algorithm on a present-day quantum computer. Specifically, we measure the Loschmidt amplitude for the Fermi-Hubbard model on a $16$-site ladder geometry (32 orbitals) on the Quantinuum H2-1 trapped-ion device. 
We assess  the effect of noise on the Loschmidt amplitude and implement algorithm-specific error mitigation techniques. By using a thus-motivated error model, we numerically analyze the influence of noise on the full operation of the quantum-classical algorithm by measuring expectation values of local observables at finite energies. Finally, we estimate the resources needed for scaling up the algorithm. 
\end{abstract}

\title{Measuring the Loschmidt amplitude for finite-energy properties of the Fermi-Hubbard model on an ion-trap quantum computer}


\author{K\'evin H\'emery}
\affiliation{Quantinuum, Leopoldstrasse 180, 80804 Munich, Germany}

\author{Khaldoon Ghanem}
\affiliation{Quantinuum, Leopoldstrasse 180, 80804 Munich, Germany}

\author{Eleanor Crane}
\affiliation{Quantinuum, Leopoldstrasse 180, 80804 Munich, Germany}
\affiliation{Joint Quantum Institute and Joint Center for Quantum Information and Computer Science,
University of Maryland and NIST, College Park, Maryland 20742, USA}

\author{Sara L. Campbell}
\affiliation{Quantinuum, 303 S Technology Ct, Broomfield, CO 80021, USA}

\author{Joan M. Dreiling}
\affiliation{Quantinuum, 303 S Technology Ct, Broomfield, CO 80021, USA}

\author{Caroline Figgatt}
\affiliation{Quantinuum, 303 S Technology Ct, Broomfield, CO 80021, USA}

\author{Cameron Foltz}
\affiliation{Quantinuum, 303 S Technology Ct, Broomfield, CO 80021, USA}

\author{John P. Gaebler}
\affiliation{Quantinuum, 303 S Technology Ct, Broomfield, CO 80021, USA}

\author{Jacob Johansen}
\affiliation{Quantinuum, 303 S Technology Ct, Broomfield, CO 80021, USA}

\author{Michael Mills}
\affiliation{Quantinuum, 303 S Technology Ct, Broomfield, CO 80021, USA}

\author{Steven A. Moses}
\affiliation{Quantinuum, 303 S Technology Ct, Broomfield, CO 80021, USA}

\author{Juan M. Pino}
\affiliation{Quantinuum, 303 S Technology Ct, Broomfield, CO 80021, USA}

\author{Anthony Ransford}
\affiliation{Quantinuum, 303 S Technology Ct, Broomfield, CO 80021, USA}

\author{Mary Rowe}
\affiliation{Quantinuum, 303 S Technology Ct, Broomfield, CO 80021, USA}

\author{Peter Siegfried}
\affiliation{Quantinuum, 303 S Technology Ct, Broomfield, CO 80021, USA}

\author{Russell P. Stutz}
\affiliation{Quantinuum, 303 S Technology Ct, Broomfield, CO 80021, USA}

\author{Henrik Dreyer}
\affiliation{Quantinuum, Leopoldstrasse 180, 80804 Munich, Germany}

\author{Alexander Schuckert}
\affiliation{Quantinuum, Leopoldstrasse 180, 80804 Munich, Germany}
\affiliation{Joint Quantum Institute and Joint Center for Quantum Information and Computer Science,
University of Maryland and NIST, College Park, Maryland 20742, USA}

\author{Ramil Nigmatullin}
\affiliation{Quantinuum, 13-15 Hills Road, CB2 1NL Cambridge, United Kingdom}

\date{\today}

\maketitle

Calculating the properties of quantum matter in equilibrium is at the heart of condensed-matter and high-energy physics as well as quantum chemistry. 
In particular, models containing interacting fermions are key to understanding high-temperature superconductivity \cite{Keimer2015} and the low-energy properties of quantum chromodynamics~\cite{PRXQuantum.4.027001}.
However, despite decades of method development, it remains challenging for classical methods to calculate equilibrium properties of high-dimensional systems with a sign problem such as spin models on frustrated lattices and fermionic models. A paradigmatic example of such systems is the two-dimensional Fermi-Hubbard model \cite{Hubbard1963, Hubbard1964}. It has attracted a tremendous amount of interest due to its rich but partially understood phase diagram \cite{Wietek2021,Schaefer2021,LeBlanc2015} and its potential application to high temperature superconductivity \cite{Timusk1999,Norman2005,Lee2006}.  

In the last ten years, analog quantum simulators have established themselves as a complementary means to studying equilibrium properties of fermions~\cite{reviewFH_Bohrdt_2021, coldatomFH_Mazurenko_2017, 256_Ebadi_2021}, although both reaching low enough energies and tuning the Hamiltonian beyond a restricted parameter regime remain challenging. Extensive progress has recently been made in the size and control of digital quantum computers, potentially leading to a highly flexible tool for solving high dimensional fermionic problems. Although many ground-state studies have been performed~\cite{vqe_tilly_2022, vqeFH_Stanisic_2022, qaoa_farhi_2014}, only few demonstrations of experimentally scalable \textit{finite-energy} or \textit{finite-temperature} quantum algorithms have been carried out so far \cite{Turro2023,Summer2023}.

Recently, \textit{time series algorithms} have been suggested as an efficient way to obtain equilibrium observables in quantum computers~\cite{lu_algorithms_2021, finite_Schuckert_2022}. These algorithms require access to only short-time dynamics - i.e. low depth circuits - on the quantum computer while the equilibrium properties are obtained by classical post-processing. 
 Despite this relative simplicity, the execution of time series algorithm on current quantum computers is still challenging due to the requirement to measure the Loschmidt amplitude
 \begin{align}
    \label{eq:loschmidt}
    G_\psi(t) = \braket{\psi| e^{-i\hat H t}|\psi},
\end{align}
where $\hat H$ is the Hamiltonian, $\ket{\psi}$ is an initial state, and $t$ is time. Indeed, the existing experimental methods for the measurement of \eqref{eq:loschmidt} require the measurement of a global observable, making them particularly susceptible to noise. 
In this work, we experimentally assess the feasibility on current quantum hardware of the quantum sub-routine of the time-series algorithm of Ref.~\cite{lu_algorithms_2021} --- the computation of the Loschmidt amplitude --- for the simulation of the Fermi-Hubbard model. To this aim, we carry out an experiment in Quantinuum's $32$ qubit digital quantum computer~\cite{Moses2023}. We analyze the effect of the noise present on the hardware, implement error mitigation strategies and extrapolate our results to evaluate the resource requirements needed to scale up the algorithm to larger system sizes. While we study the Loschmidt amplitude in the context of time series algorithm, note that the kind of interferometry experiment we performed in this work has important uses in other quantum algorithms \cite{Yang2023b}, notably in quantum phase estimation algorithms \cite{OBrien_2019,Yi2023,Lin2021,Ding2023,Somma2019}, which has been demonstrated on Quantinuum hardware on small systems with error detection \cite{Yamamoto2023}.

Our main findings are twofold. First, a quantum computer with average gate fidelity of 0.998, low state preparation and measurement (SPAM) error as well as all to all connectivity--such as the H2 device--allows for the extraction of physical properties of the Fermi-Hubbard model at finite energies using times series algorithms. Second, scaling to the classically intractable problems is expensive without further improvement due to the large shot overhead associated with error mitigation as well as the cost of performing Monte Carlo sampling.

To begin with, we summarize in Section \ref{sec:algorithm} the algorithm proposed in Ref.~\cite{lu_algorithms_2021} and the protocol we use to measure the Loschmidt amplitude. In section \ref{sec:model}, we introduce the Fermi-Hubbard model and explain how we map fermions to qubits in order to perform the dynamics on the digital quantum computer. In section \ref{sec:implementation_error_model} we measure the Loschmidt amplitude of a product initial state for the Fermi-Hubbard model on the ladder geometry, and apply error mitigation schemes to the data obtained from the quantum device. Then, in section \ref{sec:montecarlo}, we simulate the full operation of the algorithm using matrix product state simulations and test the sensitivity of the algorithm to the presence of noise, as it is unlikely that all errors can be mitigated in the near future. Finally, motivated by these results, we finally evaluate the feasibility of this algorithm and discuss its prospects for quantum advantage in section \ref{sec:Q_advantage}.

 \begin{figure*}
    \centering
    \includegraphics[width=\linewidth]{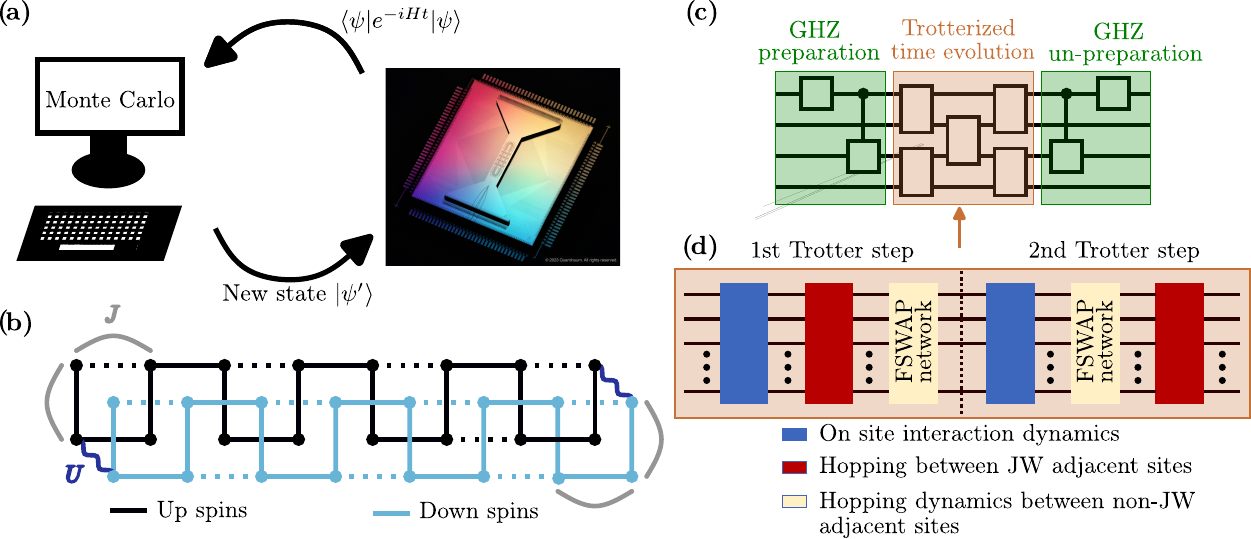}
    \caption{\textbf{Hybrid quantum-classical algorithm for finite energy properties of the Fermi-Hubbard model on the H2 quantum computer.} (a) Illustration of the quantum-classical loop in the algorithm of Ref.~\cite{lu_algorithms_2021} with a modified picture of the Quantinuum H2 surface ion trap microchip used in our work. (b) The mapping of the Fermi-Hubbard model on a $2\times 8$ lattice onto $32$ qubits. The spin-up (spin-down) fermions are encoded the black (blue) half of the system. The JW-adjacent sites are linked by a solid line, while JW non-adjacent sites are linked by a dashed line. (c) Sketch of the circuits used to measure the Loschmidt amplitude using the GHZ-like state preparation. (d) Details of the structure of the time evolution circuit using two Trotter steps. 
    }
    \label{fig:coceptual}
\end{figure*}

\section{Algorithm}
\label{sec:algorithm}

\subsection{Review of the time series algorithm for the microcanonical ensemble}
\label{sec:timeseries}
To compute observables of excited states we implement the quantum subroutine of an algorithm put forward in \cite{lu_algorithms_2021}. The underlying idea behind this algorithm is that the expectation value of an energy-filtered state with low variance will approach the micro canonical expectation value, even if the width of the filter does not tend toward zero. This can be understood in light of the eigenstate thermalization hypothesis \cite{Deutsch1991,Srednicki1994,Rigol2008} which predicts that the expectation values of few-body observables operators is a smooth function of energy, implying that they do not change abruptly within a small energy window. Although such a low energy variance state is difficult to prepare directly on the quantum computer, one can nevertheless use a cosine filter operator (see also Ref. \cite{finite_Schuckert_2022} for a different perspective akin to Wick rotation), which can be decomposed into a sum of time evolution operators. For convenience, the main steps of the calculation presented in Ref. \cite{lu_algorithms_2021} are reproduced in appendix \ref{sec:review_lbc}.   

The central quantity required for the algorithm is the Loschmidt amplitude given in Eq. \eqref{eq:loschmidt}.
We explain in the next section how to efficiently measure this quantity. 
From the Loschmidt amplitude measured at different times, one can approximately calculate the filtered density of states $D_{\psi,\delta}(E)$ (see appendix \ref{sec:review_lbc}) defined as: 
\begin{equation}
D_{\psi,\delta}(E)=\langle \psi | e^{- \frac{(\hat H-E)^2}{2 \delta^2}}| \psi \rangle.
\end{equation}
$D_{\psi,\delta}(E)$ can be understood as a weighted sum of the overlaps of $|\psi\rangle$ with the eigenstates inside a Gaussian energy filter of width $\delta$ centered around the energy $E$. The longer one performs the time evolution, the smaller the width of the filter $\delta$ becomes.
Supposing that one is interested in an observable diagonal within the $\hat Z$-product state basis $\{ |\psi_p\rangle \text{, } p=1, \dots, 2^N\}$, one can compute the microcanonical expectation value the following way:

\begin{equation}
    \langle \hat O_\delta (E) \rangle = \frac{\sum_{|\psi_p \rangle} D_{\psi_{p},\delta}(E) O_p}{\sum_{|\psi_p \rangle} D_{\psi_{p},\delta}(E)},
    \label{eq:sum_micro}
\end{equation}
 with $O_p=\langle \psi_p | \hat O|\psi_p \rangle$ the corresponding eigenvalue of $\hat O$. Instead of calculating $D_{\psi_{p},\delta}(E)$ for every $|\psi_p \rangle$ in order to evaluate the sum in equation (\ref{eq:sum_micro}), one can simply use a classical, sign problem-free Monte-Carlo algorithm to efficiently sample from the distribution, provided than one can measure the Loschmidt amplitude on a quantum device. 
 In this work, we assess the effect of the noise present on current hardware on the program outlined above and estimate the resources necessary to its application for larger system sizes.
\subsection{GHZ-like state preparation for the measurement of the Loschmidt amplitude}
The Loschmidt amplitude~(\ref{eq:loschmidt}) corresponds to the (generally complex) overlap of a time-evolved state with the initial state. This quantity can be calculated on a digital quantum computer using the Hadamard test, where the dynamics $U(t)$ of a quantum system is controlled on an ancilla qubit which starts in a superposition and interferes with both evolved and non-evolved systems when it is rotated out of the superposition~\cite{Hadamard_test}.
However this method is costly in terms of entangling gates, as it requires controlling every gate of $U(t)$ on an ancilla qubit. Alternatively, all qubits can be prepared in a GHZ-like state corresponding to a superposition between the state of interest and a state that does not evolve under application of the Hamiltonian, i.e. an eigenstate $|\phi \rangle $. The Hamiltonian is then applied and the qubits rotated back into the original basis, causing interferometry between the time-evolved and initial states~\cite{lu_algorithms_2021,finite_Schuckert_2022} -- as illustrated in Fig. \ref{fig:coceptual}. More precisely, if we define the states $|\pm (\phi,\psi_0,\xi) \rangle = \frac{1}{\sqrt{2}} \left(|\phi \rangle \pm e^{i \xi}|\psi_0 \rangle \right)$, the real part of the Loschmidt amplitude can be extracted from measurements of  $||\langle + (\phi,\psi_0, Et) | \hat U(t) | \pm (\phi,\psi_0,Et) \rangle ||^2$ in the following way: Consider the quantities 
\begin{subequations}
\label{eq:def_p0_ppi}
\begin{equation}
\begin{split}
p_0(t):&=||\langle +(\phi,\psi,Et) | \hat U(t) | + (\phi,\psi,Et)\rangle ||^2\\ 
       &= \frac{1}{4} \left (1+|G_\psi(t)|^2+2 \Re \{ G_\psi(t) e^{iEt} \} \right)
\end{split}
\end{equation}
and 
\begin{equation}
\begin{split}
p_\pi(t):&=||\langle +(\phi,\psi,Et) | \hat U(t) | - (\phi,\psi,Et)\rangle ||^2\\ 
         &= \frac{1}{4} \left (1+|G_\psi(t)|^2-2 \Re \{ G_\psi(t) e^{iEt}\}\right ).
\end{split}
\end{equation}
\end{subequations}
Therefore we find:
\begin{equation}
\Re \left ( G_\psi(t) e^{iEt} \right) = p_0(t)-p_\pi(t).
\end{equation}
We have thus reduced the problem of measuring the non-Hermitian observable $\hat U(t)$ to measuring the real quantities $p_0$ and $p_\pi$ which correspond to the probabilitity of the circuit shown in Fig.~\ref{fig:coceptual}c to output certain bitstrings.
We review the details of this technique in appendix \ref{sec:ghz}. Compared to the conditional dynamics technique outlined before, this method reduces the circuit depth by a significant factor, as highlighted in Table \ref{tab:resource_count}. Note that a GHZ-state preparation can be achieved using a constant depth circuit with mid-circuit measurement or a log-depth circuit without mid-circuit measurement. A GHZ-state preparation with 32 qubits has been carried out with $82\%$ fidelity on the device used in this study \cite{Moses2023}.

We note that in the case where the initial states are product states, it is also be possible to apply a series of single qubit interferometry experiments in order to circumvent the GHZ-state preparation \cite{lu_algorithms_2021,finite_Schuckert_2022}. However, this method introduces a shot overhead proportional to system size, and is susceptible to error accumulation. Furthermore, a new interferometry technique employing a short imaginary time evolution has been very recently introduced \cite{Yang2023b}, and could also be used for this algorithm.

\section{Model and quantum circuit implementation}
\label{sec:model}

The Hamiltonian for the Fermi-Hubbard model (FH) is given by

\begin{equation}
H = H_\text{hopp} + H_\text{int}
\end{equation}

\begin{equation}
H_\text{hopp} = -J \sum_{\langle i, j \rangle,\sigma} \left( a^\dagger_{i\sigma} a_{j\sigma} + a^\dagger_{j\sigma} a_{i\sigma}  \right),
\end{equation}

\begin{equation}
H_\text{int} = U \sum_i n_{i \uparrow} n_{i \downarrow},
\end{equation}
where $a_{i\sigma}$ ($a^\dagger_{i\sigma}$) is a fermionic operator that destroys (creates) a particle at site $i$ with spin $\sigma$, $n_{i\sigma}=a^\dagger_{i\sigma}a_{i\sigma}$ is the number operator, and $\langle i, j \rangle$ denotes adjacent sites on a lattice. The term $H_\text{hopp}$ is the hopping term of the Hamiltonian, which enables fermions to move to neighbouring sites. The term $H_\text{int}$ describes the on-site interactions between spin-$\uparrow$ and spin-$\downarrow$ fermions. The $J$ and $U$ are the parameters that control the magnitude of the hopping and interaction terms. Throughout this paper, we choose $J=0.5$ and $U=2$.

To encode the fermionic operators on a quantum computer we use the Jordan-Wigner (JW) transform, which maps each fermionic mode to one qubit such that the qubits are interpreted as lying along a 1D line. The hopping term, $H_\textrm{hopp}$, is mapped to

\begin{equation}
J (a^\dagger_{i\sigma} a_{j\sigma}+ a^\dagger_{j\sigma} a_{i\sigma}) \rightarrow \frac{J}{2} \left(X_{i\sigma} X_{j\sigma} + Y_{i\sigma} Y_{j\sigma} \right) Z_{i+1,\sigma} ... Z_{j-1,\sigma},
\end{equation}
where $X_{j\sigma}$, $Y_{j\sigma}$ and $Z_{j\sigma}$ are the Pauli operators acting on $j$th site of $\sigma \in \{\uparrow,\downarrow\}$ spin sector. The interaction term $H_\textrm{int}$ is mapped to
\begin{equation}
U n_{i\uparrow} n_{j\downarrow} = U a^\dagger_{i\uparrow} a_{i\uparrow} a^\dagger_{j\downarrow} a_{j\downarrow} \rightarrow \frac{U}{4} \left( I_{i\uparrow} - Z_{i\uparrow}\right)\otimes \left( I_{i\downarrow} - Z_{i\downarrow}\right).
\end{equation}

The sites which are adjacent in the Jordan-Wigner ordering will be referred to as JW-adjacent. 
All of the terms of the Hamiltonian between the JW-adjacent sites are two qubit operators. The terms between non-JW-adjacent sites involve Pauli strings whose length is proportional to the distance between sites in the JW orderin. For for a $2 \times 8$ rectangular lattice the interactions are illustrated in Figure \ref{fig:coceptual} (b).  

Various methods have been proposed in the literature to perform Hamiltonian simulation on a digital quantum computer, such as Trotter decomposition \cite{SUZUKI1990319}, randomly compiled Hamiltonian simulation \cite{Childs2019,Campbell2019} or classically optimized quantum simulation \cite{McKeever2022,Maurit2023,Mansuroglu2023}. We use a first order Trotter decomposition, which approximates $\hat{U}(t)=e^{-i H t}$ by

\begin{equation}
\hat{U}_{\textrm{Trotter}}(t) =  \left(e^{-i H_\text{hopp} \Delta t }e^{-i H_\text{int} \Delta t} \right)^n. \label{eq:trotterizaiton}
\end{equation}
where $\Delta t = t/n$ and $n$ is the number of steps.
For generic observables, one would expect the first-order Trotter decomposition to lead to an error $ \mathcal{O}(t^2/n)$. However, for the Loschmidt amplitude of a Hamiltonian and  initial states that are real in the same basis, the first-order decomposition turns out to be surprisingly efficient: As we show in Appendix~\ref{app:trotter_scaling}, we have 
\begin{equation}
\langle \psi | \hat{U}_{\textrm{Trotter}}(t) |\psi \rangle = \langle \psi | e^{-i\hat{H}t} |\psi \rangle + \mathcal{O}(t^3/n^2),
\end{equation}
i.e., the first order-decomposition scales just as well as the second-order one, gaining a factor $t/n$ over the naive scaling.

We now focus on the Trotter circuit implementation on the H2 quantum computer. The qubits in the H2 charge-coupled device ion trap quantum computer are effectively all-to-all connected, as any ion pair can be brought into interaction zones via shuttling \cite{Kielpinski2002, Pino2021}. The native two qubit entangling gate on H2 is the ZZphase gate, which implements an $e^{-i Z_i Z_j \theta}$ operation between two qubits $i$, $j$ with a tunable phase $\theta$. The interaction Hamiltonian $H_{\mathrm{int}}$ contains the $Z^\uparrow_i Z^\downarrow_i$ two-body terms. Its time evolution can thus be directly realized using the $N$ ZZphase gate, where $N$ is the number of sites. Note that if the initial state $\psi$ is a classical bit string state, then the effect of $e^{H_\textrm{int}}$ in the first Trotter step can be implemented using single qubit rotations.

The hopping dynamics between sites which are adjacent in the Jordan-Wigner ordering
is implemented using operators $e^{i( X X + YY) \alpha }$, where $\alpha = -J \Delta t$.  This can be expressed as a product of $\textrm{XXPhase}(\alpha)\equiv e^{i XX \alpha}$ and  $\textrm{YYPhase}(\alpha)\equiv e^{i YY \alpha}$, both of which are equivalent to ZZPhase up to a conjugation by local unitaries. Thus, the cost of implementing hopping dynamics between two JW-adjacent sites is 2 two-qubit gates.
The hopping between the non-JW adjacent sites is more complex since it involves operators with long Pauli strings, $e^{i (XX+YY)Z...Z \alpha}$.
An elegant way to compile these operators into two-qubit gates is using the fermi-SWAP (FSWAP) networks \cite{nielsen_chuang, whitfield2011}. The FSWAP gates, defined as \mbox{$\textrm{CZ}\cdot \textrm{SWAP}$}, swap the states of JW-adjacent fermions while preserving the anti-symmetric exchange symmetry of the statevector. A sequence of FSWAP gates can be used to bring the distant fermions into JW-adjacent position. Once the sites are JW-adjacent, the hopping dynamics can be implemented as usual using 2 two-qubit gates. 
The cost of implementing the FSWAP operation on H2 is only 1 two-qubit gate, since the SWAP can be implemented by simply relabelling the qubits, and the CZ gate can be implemented using $\textrm{ZZPhase}(\pi/4)$ and local rotations. One round of application of FSWAP network changes the ordering of qubits. Thus to restore the original order, the gate sequence in the next Trotter step is reversed as shown in Figure \ref{fig:coceptual}d. For a square $L\times L$ the gate overhead associated with FSWAP gates scales as $L^3\propto N^{\frac{3}{2}}$, where $N$ is the number of qubits (cf. Appendix \ref{sec:FSWAP}).

For a general $x \times y$ rectangular lattice, the total number of gates for two Trotter steps is given in Table \ref{tab:resource_count}.
Measuring $\Re \langle \psi | U(t) | \psi \rangle$ using the GHZ-state technique adds only a  linear in system size overhead to the two qubit gate count of \mbox{$n-2$} gates, where $n$ is the number of fermions in the state $|\psi \rangle$ which is significantly smaller than measuring the Loschmidt amplitude using the Hadamard test: For the 2x8 and 5x5 lattices the GHZ technique results in approximately a factor of three reduction in 2-qubit gate count. For details of the gate decomposition into the native gateset, we refer to Appendix \ref{sec:FSWAP}.

\begin{table}%
       \centering
       \subfloat[][GHZ-state technique]{\begin{tabular}{|c|l|l|l|l|}
\hline
\multicolumn{1}{|l|}{}          &                       & \multicolumn{2}{l|}{Lattice}                                            &    \\ \cline{3-5} 
\multicolumn{1}{|l|}{}          &                       & \multicolumn{1}{l|}{$x\times y$}                          & $2\times 8$ & $5\times 5$\\ \hline
\multirow{4}{*}{\makecell{\#2qb \\ gates}}    & onsite interaction   & \multicolumn{1}{l|}{$x y$}                                & 16         & 25 \\
                                &   hopping interaction             & \multicolumn{1}{l|}{$y x^2 + 7 x y-4(x+y)$}                & 104        & 260 \\
                                & GHZ preparation           & $2 (xy-1)$         & 30             &   48        \\ \cline{2-5} 
                                & 2 trotter steps     & \multicolumn{1}{l|}{$2y x^2+17 y x -8 (x+y) - 2$} & 254        & 593 \\ \hline
\multicolumn{1}{|l|}{\# qb} &                       & \multicolumn{1}{l|}{$2 x y$}                              & 32         & 50 \\ \hline
\end{tabular}}%
       \qquad
       \subfloat[][Hadamard test]{\begin{tabular}{|l|l|lll|}
\hline
                             &                    & \multicolumn{3}{l|}{Lattice}                                                                                \\ \hline
                             &                    & \multicolumn{1}{l|}{$x \times y$}                      & \multicolumn{1}{l|}{$2\times 8$} & $5\times 5$ \\ \hline
\multirow{3}{*}{\makecell{\#2qb \\ gates}} & onsite interaction & \multicolumn{1}{l|}{$7 x y$}                           & \multicolumn{1}{l|}{112}         & 175         \\
                             & hopping interaction            & \multicolumn{1}{l|}{$5 y x^2+11x y-8(y+x)$}    & \multicolumn{1}{l|}{256}         & 820         \\
                             & 2 trotter steps  & \multicolumn{1}{l|}{$29 x y+10 y x^2-16(y+x)$} & \multicolumn{1}{l|}{624}         & 1815        \\ \hline
\# qb                   &                    & \multicolumn{1}{l|}{$2 x y + 1$}                       & \multicolumn{1}{l|}{33}          & 51          \\ \hline
\end{tabular}}
      \caption{
      The two-qubit gate cost for implementing trotterized time evolution and measurement of $\textrm{Re}[\langle \psi | U_\text{Trotter} | \psi \rangle ]$ for a  $x\times y$ rectangular Fermi-Hubbard lattice using (a) GHZ-state technique and (b) the Hadamard test.  }
     \label{tab:resource_count}
     \end{table}

\section{Results on the quantum device}
\label{sec:implementation_error_model}

\begin{figure*}
	\centering
    \includegraphics{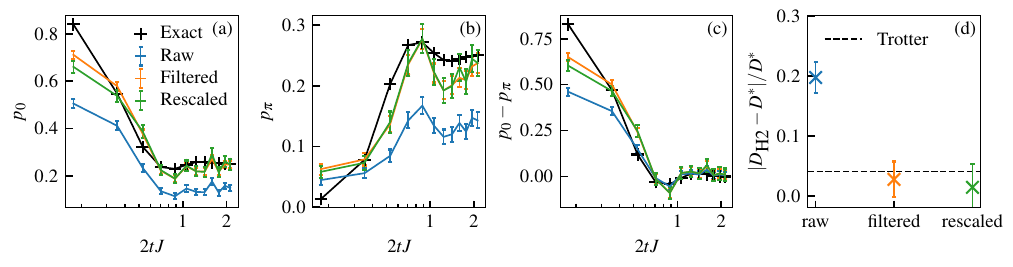}
	\caption{\textbf{Experimental data for the Loschmidt amplitude.} For each time point, we prepare a GHZ-state $|0\dots 0 \rangle + e^{iEt} |\psi_0 \rangle$ with $|\psi_0 \rangle = |1010\dots \rangle$ the Neel state on a 2x8 ladder (32 spin orbitals) and $E=1.0$. Subsequently, the system is evolved under the Fermi-Hubbard model with $J=0.5, U=2$ using Trotterised time evolution. Finally, the inverse of the GHZ-state preparation circuit (without the extra phase $e^{iEt}$) is applied and the probability of the bitstrings $p(0\dots 0) = p_0$ and $p(10 \dots 0) = p_\pi$ is measured at the output. The number of Trotter steps is set to one for the first two time-points and two for all other time points. (a), (b), (c) Results for $p_0= \frac{1}{4} || 1+e^{iEt} G_{\psi_0}(t)||^2$,  $p_\pi= \frac{1}{4} || 1-e^{iEt} G_{\psi_0}(t)||^2$ and $p_0-p_\pi=\Re(e^{iEt} G_{\psi_0}(t))$ compared with exact classically simulated circuits and with data obtained by applying the error mitigation techniques described in section~\ref{sec:implementation_error_model}. Note that the raw data for $p_\pi$ is closer to the exact result than the mitigated data for the first time point. (d) Comparison of the filtered density of states obtained from the exact time evolution $D^{*}$ with 8 Trotter steps with the one obtained from the experimental estimates $D_{H2}$. The filtered density of states are obtained by using $D_{\delta,|\psi\rangle}(E) \approx \sum_{m=0}^{R} 2 c_m  \Re \langle \psi | e^{-i (H-E) t_m}| \psi \rangle$ with $c_m=\frac{1}{2^M} {M \choose M/2-m}$, $\delta=1.0$ leading to $R =12$, see Appendix~\ref{sec:review_lbc}. The dotted line represents the relative error of the filtered density of states due to Trotterisation. The effect of increasing the final time and sampling rates is negligible see appendix~\ref{sec:cutoff}. 
  \label{fig1}}
\end{figure*}

\begin{figure}
	\centering
    \includegraphics{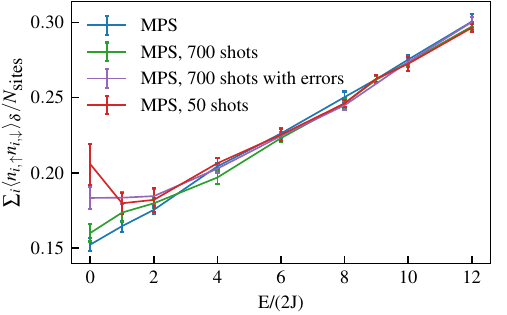}
	\caption{\textbf{Performance of the algorithm for different energies as a function of noise.} We compare the results obtained for the average double occupancy per site $\sum_i n_{i, \uparrow} n_{i, \downarrow}/N_\text{sites}$ from the Markov chain obtained by simulating the Loschmidt amplitudees using matrix product state (MPS) time evolution (blue) to results obtained by adding shot noise for 700 shots (green) and adding shot noise and Gaussian noise (red). For all data points considered, we used 5000 Monte Carlo samples, except for $E/(2J)=0.0, 1.0 \textrm{ and } 2.0$ in the presence of error, for which we used $8800$ samples. The error bars come from a blocking analysis of the Monte Carlo samples \cite{Flyvbjerg1989}. For all the simulations, the bond dimension used was $\chi=100$.}
 \label{fig:montecarlo}
\end{figure}

In order to test the methods discussed above, we benchmark the measurement of the Loschmidt amplitude for $2\times 8$ FH model on the 32 qubit H2 Quantinuum device. As the initial state, we choose the N\'eel state: $|\psi_0 \rangle =a_{1,\uparrow}^{\dagger}a_{2,\downarrow}^{\dagger} a_{3,\uparrow}^{\dagger} \dots |0 \rangle$, where the ordering of sites follows a ``snake" as illustrated on Fig. \ref{fig:coceptual}. The results are shown in Fig.~\ref{fig1}. Since the Loschmidt amplitude is a global observable, a single error in any of the gates will cause a corruption of the output. Therefore, we would generically expect to measure a signal that is reduced from its ideal value by a factor
\begin{equation}
\label{eq_qfactor}
    q = \prod_{i} F_i
\end{equation}
where $F_i$ is the fidelity of gate $i$ and the product runs over all gates in the circuit. In our case, assuming all gates to be of the same quality, we expect \cite{Yang2023} 
\begin{align}
\Re \, G^\mathrm{noisy} = q \Re \,  G^\mathrm{noiseless}
\label{eq_qshift}
\end{align}
with $q \approx 99.8\%^{n_\mathrm{two-qubit \ gates}}$, 
on average \cite{Moses2023}. Thus, the most straightforward error mitigation scheme is to simply \textit{rescale} the obtained results according to the above formula.
Alternatively, we compare this method to \textit{symmetry-filtered post-processing}. As our trotterized time evolution conserves the number of spin-up and spin-down fermions in the system, we can discard the shots where the bit-strings do not correspond to the initial number of particle \cite{vqeFH_Stanisic_2022}. 
The data shown in Fig.~\ref{fig1} demonstrates that both techniques yield similar results.

While the rescaling by a factor equal to the inverse of the global fidelity works reasonably well, there are corrections to this simple rescaling. Surprisingly, the raw signal obtained for the first time point for $p_{\pi}$ is greater than the clean value, which can not be captured by the model given by Eq. \eqref{eq_qshift}. The reason for this counterintuitive result is that the GHZ state that is involved in time is highly non-generic: incoherent $Z$-errors cause a flip between the $| (\phi,\psi,Et)\rangle$ and $|- (\phi,\psi,Et)\rangle$ states (cf. sections \ref{sec:error_mitigation} and \ref{sec:memory_error}), which can \textit{increase} the values of the measured probabilities $p_0$ and $p_\pi$. 
Furthermore, we expect memory errors to play a larger role with increasing system size. These errors can be modeled as coherent evolution $e^{i\sum_i Z_i\theta_i}$ where $\theta_i$ are angles that depend on the idling time of qubit $i$. In particular, a translationally invariant memory error maps $\Re G \rightarrow  \cos(\sum \theta_i) \Re G + \sin(\sum \theta_i) \Im G$ and thus $p_0$ or $p_\pi$ can be larger than their noiseless values.
Mitigating the effect of coherent errors on the measurement of the Loschmidt amplitude is beyond the scope of the present work, but clearly important to explore in the future, for example, by incorporating dynamical decoupling techniques \cite{Viola1999,Smith2021}.

We note that the rescaling error-mitigation method requires a good device characterisation as its results will be only as precise as the knowledge of the gate fidelity. In contrast, the symmetry-filtering method is device-agnostic and does not depend on any calibration parameter. Both methods come at the cost of increasing the uncertainty by a factor $q^{-2}$, which can in turn be offset by increasing the number of shots by the same factor. As we will now see, the estimate yielded by both error mitigation techniques is sufficiently close to $G^\mathrm{noisesless}$ for the target parameters.


\section{Classical simulation of the Monte Carlo sampling}
\label{sec:montecarlo}
In order to assess the effect of the noise on the final expectation value of the operator of interest in the microcanonical ensemble using classical simulations, we assume that all initial states behave similarly in presence of noise. We simulate the Markov chains at different energies for the ladder geometry using  MPS techniques, with bond dimension $\chi=100$. While it is not expected that the precise behaviour of the Loschmidt amplitude at the longest times are perfectly captured with this bond dimension, the filtered densities of states nevertheless converges quickly with bond dimension (see appendix \ref{sec:bond_dimension}), in line with the findings of Ref. \cite{finite_Schuckert_2022,Yang2022}. Therefore, the bond-dimension captures the correct features of the target (unormalized) distribution of the sampling algorithm.
To simulate the effect of shot noise of ideal quantum hardware, we add binomial noise to the time series.
As can be seen in Fig. \ref{fig1}, some appreciable bias beyond shot noise remains which after the various error mitigation procedures. Therefore, to simulate the effect of systematic hardware errors that can not be perfectly mitigated, we add random error terms to the time-series. They are drawn from a Gaussian distribution centered at zero with standard deviation $\sigma=0.05$. With this simple error-model, we artificially introduce more error than observed on Fig. \ref{fig1}.   
In each case, we use the noisy timeseries to calculate the filtered density of states. The results are shown on Fig. \ref{fig:montecarlo}.

A few comments are in order. First, the quantity represented in Fig. \ref{fig:montecarlo}: $\sum_i \langle n_{i,\uparrow}n_{i,\downarrow}\rangle_{\delta}/ N_\textrm{sites}$, with $\delta=1$, is the expectation value of the double occupancy per site in the \textit{filter ensemble}. For finite systems, the filter ensemble can be thought of as a moving average of the micro-canonical ensemble expectation value over an energy window of width $\delta$. By keeping $\delta$ of the order $O(1)$ and by increasing system size, the filter ensemble eventually converges to the microcanonical ensemble for intensive quantities and for generic quantum systems which satisfy the eigenstate thermalization hypothesis \cite{lu_algorithms_2021, Yang2022}. However, Fig. \ref{fig:montecarlo} already captures the tendency of the Fermi-Hubbard model to be insulating at high energy and conducting at low energies.
 Second, the finite-energy algorithm in presence of noise displays a similar behaviour to the finite-temperature scheme investigated in Ref. \cite{Ghanem2023}. Namely, the expectation values of local observables are not sensitive to noise at high energies/temperatures. The algorithm starts to show deviations to the noiseless values close to $E=0.0$. Note that $E=0.0$ is the lowest energy that can be targeted in a scalable manner with $\hat Z$-product states. As explained in Ref. \cite{lu_algorithms_2021,Yang2022}, it is not possible to explore lower energies using the $\hat Z$-product state basis, since the low overlap of these initial states with the corresponding eigenstates will decay with system size, yielding a vanishing density of states when approaching the thermodynamic limit. In contrast with the finite-temperature algorithm, we find that we need very few shots (as low as 50 for the energy considered) to converge towards the correct expectation value. This indicates that the finite-energy scheme is much more resilient to noise than the finite-temperature algorithm. This is explained by the fact that the Boltzman weights $W_\psi$ are not simply equal to the density of states as in the present work, but are convolved by a factor $e^{-\beta \omega}$: $W_\psi= \int dE e^{-\beta E} D_\psi(E)$. Therefore, the low energy sector is multiplied by an exponential factor, and any error at low energy caused by noise will be amplified accordingly. Our results thus support the hypothesis that finite-energy properties are more amenable to quantum techniques in the near-term than finite-temperature ones, at least in the time series framework proposed in~\cite{lu_algorithms_2021}.
However, we note that other initial states---with higher overlap with the low-energy sector---could be chosen, as demonstrated in Ref. \cite{Yang2022}. It would be interesting to compare the performances of both schemes when sampling from these initial states.
\section{Prospects of Quantum Advantage} 
\label{sec:Q_advantage}
\begin{figure}
    \centering
    \includegraphics{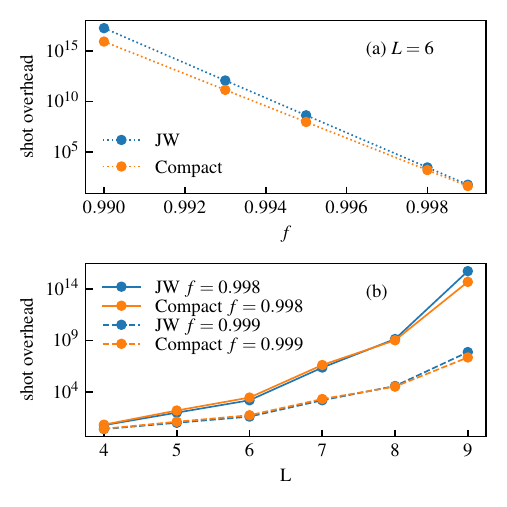}
    \caption{\textbf{Comparison of the estimated shot overhead between Fermion encodings.} We consider the cost of time evolving a  square lattice Fermi-Hubbard model of size $L \times L$ with $J=0.5, U=2$ up to $t=1$ with a Trotter error comparable to the one demonstrated in the experimental results of this work. The shot overhead $1/q^2$ is the number of shots required to compensate the exponential damping of the signal by the global fidelity, given in terms of the  fidelity to the power of the number of two-qubit gates $q$, cf. Eq.~(\ref{eq_qfactor}). We assume that single-qubit gates can be carried out with unit fidelity. We compare the Jordan-Wigner (JW) encoding and the compact encoding for two-qubit gate fidelity $f=0.998$ and $f=0.999$. A second-order Trotterisation is assumed for all time points which, in order to reach the same precision as in the present experiment, requires two Trotter steps at $L=4$, three steps at $L=5,6$ and four steps at $L=7,8$.} 
     \label{fig:ressource_estimate}
\end{figure}
It has been demonstrated in Ref. \cite{Yang2022} that, for the finite-energy algorithm studied in our work, choosing the maximal time constant as a function of system size  is sufficient to reach the microcanonical expectation value in the thermodynamic limit, for generic quantum systems satisfying the eigenstate thermalization hypothesis. Naively, one might conclude that quantum circuits of constant depth are thus sufficient. 
However, as the system size increases, so does the Trotter error. It has been proven in Ref. \cite{Childs2021} that for a local Hamiltonian acting on $N$ qubits the number of Trotter steps required to reach time $t$ with a fixed precision scales as $N^{1/p} t^{1+1/p}$ in the worst case with $p$ the order of the Trotter product formula. The number of entangling gates needed per second order Trotter step using the Jordan-Wigner encoding for a rectangular geometry of size $x \times y$ is given by:
\begin{equation}
n_{\text{JW}}= 2 y x^2 + 3 x y + 14 x -2 y -15
\end{equation}
while the number of entangling gates needed for a recently proposed local encoding \cite{Derby_compact_2021} is given by:
\begin{equation}
n_{\text{compact}}= 26 x y - 24(x + y)
\end{equation}
While the overall scaling of the algorithm is extremely favorable in a fault tolerant setup, in the NISQ era the strength of the signal decreases exponentially with the number of gates, as $G_\psi(E)_\text{measured} \propto f^{N_\text{gates}}:= q$, where $f$ is the gate fidelity. Therefore, the number of shots will increase exponentially with system size, in order to reach the precision that would be obtained on a noiseless, ideal device. Based on these considerations, we present the shot overhead as a function of the size of the system for the square lattice geometry in Fig. \ref{fig:ressource_estimate}. We choose the units such that $t=1$, yielding $n_\text{trotter}=\left \lceil \frac{2}{\sqrt{32}} \sqrt{N_{qubits}} \right \rceil$, such that for 32 qubits we use 2 Trotter steps as in the present work, and use $p=2$ (second order Trotter decomposition). Our resource estimate is likely pessimistic, as it would be in principle possible to take the final time as small as $t \propto 1/\sqrt{N}$ \cite{Yang2022}. However, choosing the maximum time constant ensures a faster convergence to the microcanonical value as a function of system size.  

Note that this resource estimate will likely be further improved by both hardware and software improvements. In particular, improved gate fidelity and optimization of the time-evolution circuits \cite{McKeever2023, Mansuroglu2023,Maurit2023, Astrakhantsev2022} have the potential to reduce the resources requirement by several order of magnitudes.

The whole algorithm requires at least $10^3$ Monte Carlo iterations for each energy density. For each iteration, we need about $10$ time steps, measured at least $10^2$ times each. Assuming $99.8 \%$ gate fidelity and an $6\times 6$ lattice, the shot overhead from error mitigation is about $10^2$. Therefore, each MC iteration requires about $10^5$ shots.
Assuming a shot time of around a second, the full algorithm would require a run time of more than 1000 days. 

While classical simulations of this algorithm can be performed efficiently in one dimension \cite{Yang2022} (we exploit that fact in Sec. \ref{sec:montecarlo}), MPS simulations would require computational resources growing exponentially with one of the dimension of the system.
Nevertheless, we expect that the particular quantum routine demonstrated in this work could also be performed for larger system sizes by other state of the art classical techniques.  First, two-dimensional tensor network classes could be used in principle, including Projected Entangled Pair States (PEPS). There, it is expected that computing the Loschmidt amplitude would be challenging, due to the complexity of the PEPS contraction \cite{SnehRai2023,Schuch2007,Gonzalez-Garcia2023,Vasseur2019}. 
Furthermore, neural network simulations have proven competitive for performing the dynamics of two-dimensional systems for short times \cite{Heyl2020,Carleo2017,Gutierrez2022}, and it would be interesting to investigate whether they are able to capture the Loschmidt amplitude with the precision required by the algorithm used in this work. Our investigation of the effect of noise in section \ref{sec:montecarlo} is encouraging, as it suggests that even approximates time-series could yield satisfying observable expectation values.

Nonetheless, it is desirable to go beyond the program we applied in the present work. First, in order to reach smaller energy/temperatures, one needs to prepare an initial state with significant overlap with the low energy sector, which would significantly increase both the classical and quantum resources needed \cite{Yang2022}. Furthermore, while we studied only static properties, a similar approach could give access to the finite-energy expectation values of dynamical observable, at the price of deeper circuits. Due to the exponential resources needed to perform time evolution classically with most commonly used methods \cite{Lin2022}, it is likely that such a program would be out of reach for classical computers. 


\section{Discussion and Outlook}
In this work, we demonstrated that the current capabilities of the Quantinuum H2 trapped-ion quantum device allow for the execution of the quantum subroutine of one of the simplest time-series algorithms on a condensed matter system. Although the noise of the machine still affects the results, the high fidelity of the gates as well as the low memory and SPAM error allow us to get satisfying data after error mitigation, in the sense that the physical features of the system should be well captured at the end of the hybrid quantum classical algorithm (see Fig. \ref{fig:montecarlo}). We compared two different and independent error mitigation techniques, one based on symmetry, the other based on the probability of success of our circuit, and found that both give comparable results. Furthermore, using classical simulations, we provided numerical evidence that the remaining errors, not correctly taken into account by our error mitigation schemes, would have a low impact on the final prediction of the finite-energy properties of the system. In other words, the particular Monte Carlo sampling explored here seems to be relatively resilient to noise. 

Overall, we demonstrated that while time-series algorithms require the measurement of a global observable, which is in principle maximally sensitive to noise, the precision of an existing device today is sufficient to run these hybrid quantum-classical schemes. We found that the effect of noise is not entirely explained by a global damping of the signal by the global fidelity. On the other hand, our resource estimates indicate that sampling over tens of thousands of initial states makes this algorithm prohibitively expensive to run on ion trap devices before a significant drop in cost per sample, for example caused by the advent of a manufacturing age in which many quantum computers can execute coherent evolutions of intermediate depth in parallel.

It is as yet an open question to evaluate how precisely classical methods are able capture the Loschmidt amplitude at  moderately short times for two-dimensional systems, although some theoretical studies have already been performed \cite{Wild2023}, which leaves open the possibility that the algorithm studied in this work could be carried out classically. However, more involved versions of this algorithm would be necessary to explore the low energy properties as well as the linear response behaviour of strongly-correlated systems. These would likely be very challenging to execute classically, indicating the possibility of near-term useful quantum advantage with time series algorithms.
\vspace{5pt}

\section*{Appendix \label{methods}}
\section*{Data availability}

The numerical data that support the findings of this study are available at \url{https://doi.org/10.5281/zenodo.8330634}.

\section*{Code availability}
The code used for numerical simulations is available from the corresponding author upon reasonable request.

\section*{Acknowledgements}
KH and KG are supported by the
German Federal Ministry of Education and Research
(BMBF) through the project EQUAHUMO (grant number 13N16069) within the funding program quantum
technologies - from basic research to market. AS and EC acknowledge support from the U.S. Department of Energy,
Office of Science, National Quantum Information Science Research
Centers, Quantum Systems Accelerator.



\begin{appendix}
\section{Review of the algorithm}
\label{sec:review_lbc}

The central quantity used in Ref. \cite{lu_algorithms_2021} is the Filter operator:
\begin{equation}
\hat P_\delta(E)= \exp \left ( - \frac{(\hat H-E)^2}{2\delta^2} \right),
\end{equation}
where $E$ is the target energy and $\delta$ is the width of the filter. It can be showed that, as long as $||H-E||_\infty<N \pi/2$, where $|| \cdot ||_\infty$ denotes the operator norm:
\begin{equation}
\hat P_\delta(E) \approx \cos \left (\frac{\hat H-E}{N} \right)^{\left \lfloor \frac{N^2}{\delta^2} \right \rfloor  _2}
\end{equation}
where $N$ is the system size and $\lfloor \cdot \rfloor_2$ denotes the nearest even integer. By writing the cosine as a sum of two complex exponentials, by using the binomial formula and by truncating    the resulting series, one finds \cite{lu_algorithms_2021}:
\begin{equation}
\hat P_\delta(E) \approx \sum_{-R}^{R} c_m e^{-i (H-E) t_m}
\label{eq:series}
\end{equation}
with $c_m=\frac{1}{2^M} {M \choose M/2-m}$ , $R = \lfloor x/\delta \rfloor$, $x$ being a scalar controlling the truncation of the series and $t_m=2m/N$. Furthermore, it has been shown that one can further reduce the number of measurements by choosing: $R= \lfloor x \alpha/\delta \rfloor$, $t_m=2m/\alpha$  and $\alpha \propto \sqrt{N}$.
In order to relate the microcanonical expectation value and the cosine filter operator, Ref. \cite{lu_algorithms_2021} considers:

\begin{align}
\langle \hat A \rangle_\delta(E) &= \frac{\text{tr} \hat A P_\delta(E)}{\text{tr} \hat P_\delta(E)} \\
                                 &= \frac{\sum_{i} D_{\delta,|i\rangle}(E) A_{\delta,|i\rangle}(E)}{\sum_i D_{\delta,|i\rangle(E)}},
                                 \label{sum to be sampled}
\end{align}
where the sum over the index $i$ denotes the sum over the product states in the $\hat Z$-basis and 
\begin{align}
A_{\delta,|\psi\rangle} &=\frac{\langle \psi| \hat A \hat P_\delta(E) + \hat P_\delta (E) \hat A | \psi \rangle}{2 \langle\psi| \hat P_\delta(E)| \psi \rangle}\\
D_{\delta,|\psi\rangle}(E)&=\langle \psi | P_\delta (E) | \psi \rangle.\\
                          & \approx \sum_{0}^{R} 2 c_m  \Re (\langle \psi | e^{-i (H-E) t_m}| \psi \rangle) \label{eq:real_part}.
\end{align}
The sum in Eq. \eqref{sum to be sampled} is sampled using classical Monte Carlo, with $D_{\delta,|i\rangle}$ being the (unnormalized) target distribution of the sampling algorithm. Note that in this work, we choose the observable $\hat A$ to be diagonal in the $\hat Z$-product state basis, therefore $A_{\delta,|i \rangle}$ reduces to the eigenvalue of $\hat A$ corresponding to the product state $|i \rangle$.

Furthermore, Ref. \cite{Yang2022} demonstrated that to converge to the thermodynamic limit expectation value it is sufficient to take $\delta \propto \sqrt{N}$, where $N$ is the system size, but that taking $\delta$ constant ensures a faster convergence to the microcanonical value when increasing system size. Here we choose $\alpha= 2\sqrt{N}$, $\delta=1$ and $x=1$. 

\section{Details of the Monte Carlo sampling}

In order to extract expectation values of observables from the time series algorithm outlined in the previous section, we sample equation \eqref{sum to be sampled} using Metropolis-Hasting algorithm.
As our goal is to probe the sector with a specific number of spin-up and spin down fermions, we use the following update scheme.
We propose a new state $|\psi'\rangle$ from the previous one $|\psi\rangle$ by applying a random hopping  of one fermion from one site to one of its nearest neighbouring sites unoccupied with the same spin. We then accept this new state with the acceptance ratio:
\begin{equation}
    A = \text{min}\left(1, \frac{D_{\delta,|\psi'\rangle}(E)}{D_{\delta,|\psi\rangle}(E)}\frac{P_{\psi \rightarrow \psi'}}{P_{\psi' \rightarrow \psi}} \right),
\end{equation}
where $P_{\psi' \rightarrow \psi}$ is the probability to hop from $\psi'$ to $\psi$ and is related to the number of unoccupied neighbouring sites. 
At the end of the sampling procedure, we obtain a list of product states. Since the observable we study in this work, the double occupancy, is diagonal in the product state basis, the expectation value is estimated as the average double occupancy of the sampled product states. 
\section{GHZ-like state preparation} \label{sec:ghz_technique}
\begin{figure}
\includegraphics{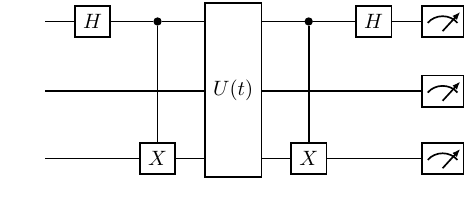}
\caption{Circuit diagram illustrating the GHZ-like state technique on 3 qubits, with a initial state $|\psi_0\rangle = |101\rangle$.}
\label{fig:cat_state}
\end{figure}

Suppose that there exists a state $| \phi \rangle$ such that $\hat U |\phi \rangle= |\phi\rangle$. Let us define $|\pm (\phi,\psi,t)\rangle = \frac{1}{\sqrt{2}} \left(|\phi \rangle \pm e^{iEt} |\psi \rangle \right)$.
We further introduce: 
\begin{equation}
\begin{split}
p_0(t)&=||\langle +(\phi,\psi,t) | \hat U(t) | + (\phi,\psi,t)\rangle ||^2\\ 
    &= \frac{1}{4} \left (1+|G(t)|^2+2 \Re \{ G(t) e^{iEt} \} \right)
\end{split}
\end{equation}
and 
\begin{equation}
\begin{split}
p_\pi(t)&=||\langle +(\phi,\psi,t) | \hat U(t) | - (\phi,\psi,t)\rangle ||^2\\ 
        &= \frac{1}{4} \left (1+|G(t)|^2-2 \Re \{ G(t) e^{iEt} \}\right ),
\end{split}
\end{equation}
where $G(t)=\langle \psi|e^{-iHt} | \psi \rangle$ as in the main text.
In order to run the microcanonical algorithm, one only needs the real part of the Loschmidt amplitude modulated by a time dependant phase, as made explicit in Eq. \eqref{eq:real_part}. It is straightforward to see that 
\begin{equation}
\Re(\langle \psi| \hat U(t) | \psi\rangle e^{iEt}) = p_0(t)-p_\pi(t).
\end{equation}
When $|\psi \rangle$ is a product state, as it is the case in this paper, this procedure is very similar to a GHZ-state preparation. Note that both $p_0$ and $p_\pi$ can be obtained from measuring all qubits of only one circuit if $|\psi \rangle$ is a product state. This can be understood by inspecting Fig. \ref{fig:cat_state}, which shows the circuit for the 3-qubits GHZ-like state with the initial product state $|\psi\rangle = |101\rangle$.  

\label{sec:ghz}
Indeed, $p_0=\frac{1}{4}||(\langle 101 |+\langle 000 |) \hat U(t) (|000\rangle+| 101 \rangle) ||^2$ correspond to the probability of obtaining the bit-string ``000''. $p_\pi=\frac{1}{4} ||(\langle 000 |-\langle 101 |) \hat U(t) (|000\rangle +| 101 \rangle )||^2$ correspond to the probability of measuring the bit-string ``000'' after introducing a single qubit $X$-gate on the right of the left-most Hadamard gate. Equivalently, $p_\pi$ corresponds to the probability of measuring the bit-string ``100''. For this reason,we denote ``000'' (resp. ``100'') the $0$-string (resp. the $\pi$-string).

We note that as an alternative to applying the inverse of the GHZ-state preparation at the end of the circuit, one may directly measure $\Pi_0 \otimes 1/M \sum_{k=1}^M (-1)^k (\cos(k \pi/M) X  + \sin(k \pi / M) Y)^{\otimes M}$, where $M$ is the number of $\ket{1}$s in the initial state and $\Pi_0$ is the projector on $\ket{0}$ on the other qubits \cite{Guehne2007}.

\begin{figure*}
    \centering
     \includegraphics[scale=0.8]{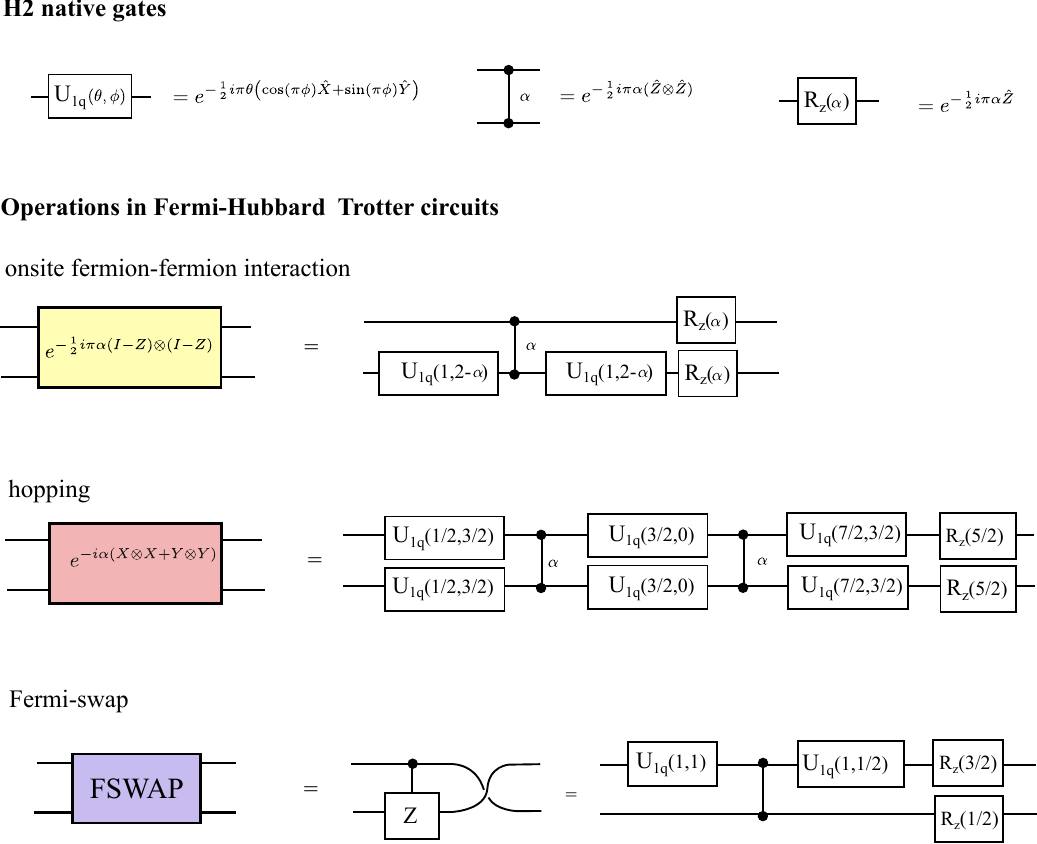}
    \caption{Operations used in the Fermi-Hubbard trotterized dynamics circuits - onsite interaction dynamics, hopping and FSWAP - and their decomposition into one- and two-qubit gates native to H2 device.}
    \label{fig:gate_decompositions}
\end{figure*}

\begin{figure*}
    \centering
      \includegraphics[scale=0.65]{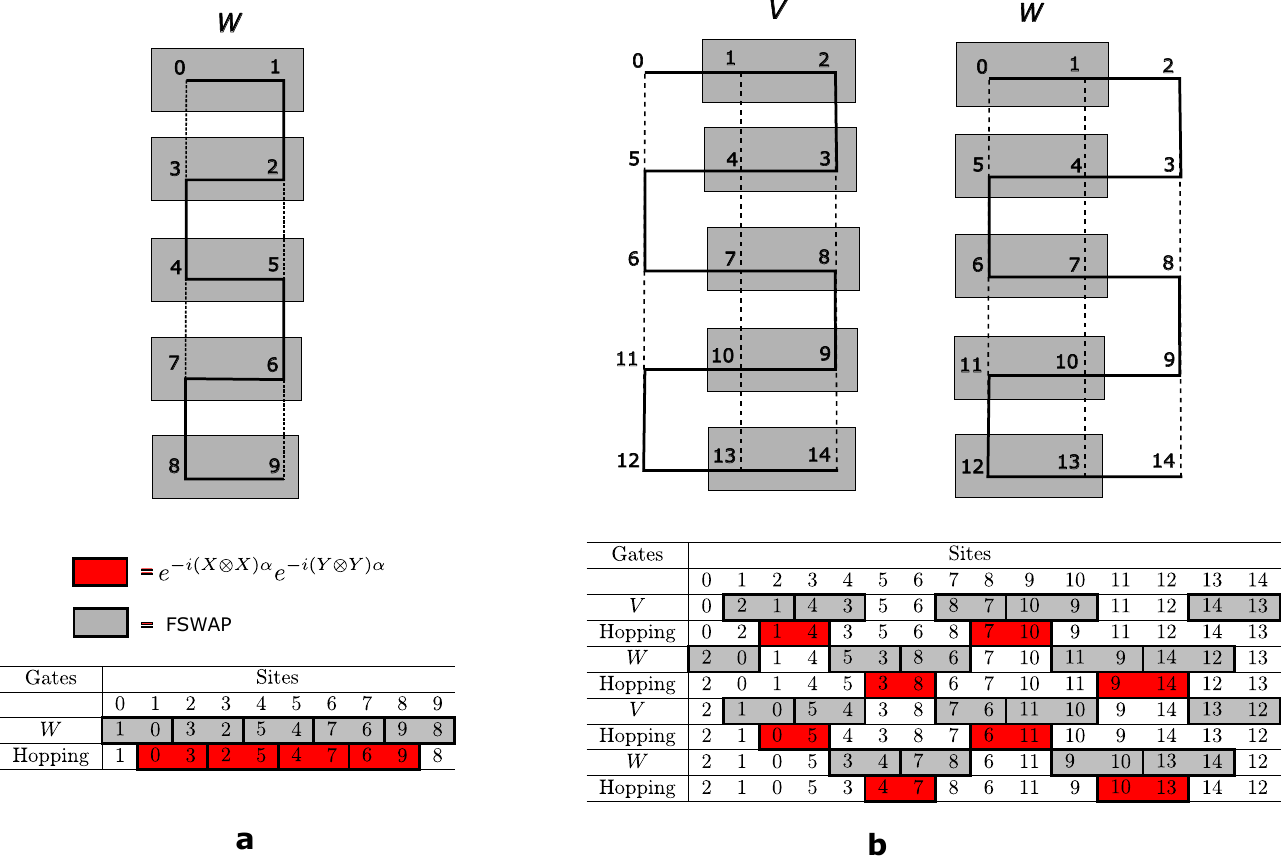}
    \caption{Fermi-swap network for $x \times y$ lattice with $x = 2$, $y = 5$ (a) and $x = 3$, $y = 5$ (b). The network consists of applying column swap operations $V$ and $W$. The tables show the full sequence of operations without returning qubits to their original order.}
    \label{fig:fswap_network}
\end{figure*}

\section{Implementing Fermi-Hubbard model Trotter circuits} \label{sec:FSWAP}

In this section we give more details on the implementation of trotterized dynamics on the H2 32 qubit quantum device.

The native one-qubit gates on the H2 are rotations $U_{1q}(\theta,\phi)\equiv e^{-\frac{1}{2}(\cos(\phi)\hat X +\sin(\phi)\hat{Y})}$ and $R_z(\theta)\equiv e^{-\frac{1}{2}i\theta \hat Z }$ for $\theta,\phi \in [0,2\pi]$, and the native two-qubit gate is ZZPhase gate implementing operation $e^{-\frac{1}{2}i\alpha (\hat Z \otimes \hat Z)}$ for $\alpha \in [0,2\pi]$ 

In the Jordan-Wigner encoding some interaction terms become strings of Pauli operators whose length is proportional to the size of the system. In the two dimensional Fermi-Hubbard model, these are the hopping terms between sites that are not adjacent in the Jordan-Wigner ordering. Operators of the form $e^{i\alpha (XX+YY)Z...Z}$ can be implemented using staircase circuit using $2(n-1)$ two qubit gates \cite{nielsen_chuang, whitfield2011}. The gate overhead associated with implementation of long Pauli strings can be reduced by using fermi-swap (FSWAP) networks \cite{fermi_swap_networks_kivlichan,phasecraft_fswap_network}. The FSWAP operator swaps the states of the neighbouring qubits in the JW ordering while preserving the fermionic anti-symmetric exchange statistics. The SWAP network is a sequence of FSWAP gates which brings the non-JW adjacent sites into a JW adjacent positions, so that the hopping term between them can be implemented locally. This can be viewed as a succession of rotations into a basis, where the non-local terms Pauli string become two body terms. 

For a rectangular grid, an efficient way to implement an FSWAP network is described in \cite{phasecraft_fswap_network}. The procedure consists of repeatedly applying the operator $V W$, where $V$ swaps odd-numbered columns with those to their right, and $W$ swaps even-numbered columns with those to their right. After each application of $V W$, a new set of qubits that were previously not JW-adjacent are made JW-adjacent, and the hopping term can be implemented locally via gate $e^{-i(X\otimes)X \alpha}e^{-i(Y\otimes)Y \alpha}$.  After implementing all of the vertical hopping interactions the fermi-swap operations would normally be applied in reverse to return the qubits into their original position. However, for trotterized dynamics this is not necessary since the order can be restored in the next trotter step, by implementing the hopping interaction gates in the reverse order. 

If the trotter circuit involves an odd number of steps then the final ordering needs to be restored by adding the fermi-swap gates in reverse at the end of the circuit.
However, for classical input states $|\psi \rangle$ (tensor products of $|0\rangle$ and $|1 \rangle$) the effect of fermi-swap network can be be efficiently computed classically and thus observable of the form $|\langle \psi|U(t)|\psi \rangle|^2$, can still be obtained without applying the reversed fermi-swap network on the last trotter step.

On H2 device, the simplest way to implement FSWAP is to use a single CZ gates and a software swap i.e. virtually relabelling the qubits. The FSWAP operator can be expressed as a product of CZ and SWAP gates, $\textrm{FSWAP=CZ} \cdot \textrm{SWAP}$.  Since H2 has all-to-all connectivity the relabelling of the qubits does not add any overheads in implementation of subsequent gates. Thus, each FSWAP operations costs one two qubit gate on H2.

The FSWAP network gate sequence for the ladder geometry and a three column geometries are illustrated in Figure  \ref{fig:fswap_network} (a) and (b) respectively. The experiments on H2 devices were carried out on a $2\times 8$ ladder geometry. In general for a $x \times y$ rectangular lattice, the number of $VW$ repetitions in the FSWAP network is $\lfloor(x-1)/2\rfloor$. Each column swap operator involves $y$ swaps and there are $(\lfloor x/2\rfloor)$ column swaps in operators $V$ and $W$. Thus, the total number of FSWAP operations in one trotter step is $y x(x-1)$. For a square 2d system, where $x=y=L$, the number of two qubit gates in FSWAP network is proportional to $\propto L^3$ or $\propto N^{3/2}$, where $N$ is the number of qubits. The superlinear scaling with the system size can be avoided by using the local fermion to qubits encoding \cite{Verstraete_2005,derby_klassen,BRAVYI2002210,Bravyi2019} instead of JW encoding, but for small systems considered in this work the JW encoding is more resource efficient.

\section{Error mitigation}
\label{sec:error_mitigation}

Error mitigation is crucial for obtaining meaningful result on NISQ devices. We have considered two different error mitigation strategies. The first strategy utilizes the number conservation symmetry of the Fermi-Hubbard trotterized evolution. It simply involves discarding the shots that violate the number symmetry and thereby reducing the error in the observed quantities. 
The second strategy involves rescaling the measured quantities to compensate for the effect of noise. Here, we detail the theoretical details of this heuristic.

The prepared GHZ-like states that undergo the trotterized evolution are given by

\begin{equation}
| \psi_0 \rangle = V_0 |0 \rangle = \frac{1}{\sqrt{2}} \left(|0 \rangle + | \psi \rangle \right),
\end{equation}

\begin{equation}
| \psi_\pi \rangle = V_\pi |0 \rangle = \frac{1}{\sqrt{2}} \left(|0 \rangle - | \psi \rangle \right),
\end{equation}
where $|0\rangle \equiv |0\rangle^{\otimes n}$ is the vacuum state and $| \psi \rangle$ is a product state. The operators $V_0$ and $V_j$ prepare the GHZ-like state from a product state. As explained in Appendix \ref{sec:ghz_technique}, $V_0$ can be constructed using a Hadamard gate on a selected qubit, $j$ and a series of CNOT gates with a control on the $j$ and targets on qubits where $| \psi \rangle$ is in state $|1\rangle$. It is easy to show that $V_\pi=X_j V_0$.

The GHZ-like measurement technique obtains real part of the Loschmidt amplitude through a difference of expectation values

\begin{align}
    \textrm{Re}[G(t)] & = |\langle 0 | V_0^\dagger U V_0 | 0 \rangle|^2 - |\langle 0 | V_\pi^\dagger U V_0 | 0 \rangle|^2 \\ 
    & \equiv |\langle 0 | W_0 | 0 \rangle|^2 - |\langle 0 | W_\pi | 0 \rangle|^2 \\  & = p_0 - p_\pi,
\end{align}
where $W_0 \equiv  V^{\dagger}_0 U V_0$ and $W_\pi \equiv  V^{\dagger}_\pi U V_0= X_j W_0$.

Now let us consider the effect of gates affected by incoherent noise on $\langle \psi_0 |U | \psi_0 \rangle$ and $\langle \psi_0 |U | \psi_\pi \rangle$. Firstly, note that a Z-flip on a single qubit can flip the GHZ-state $| \psi_0 \rangle \leftrightarrow | \psi_\pi \rangle $

\begin{equation}
Z_k \frac{1}{\sqrt{2}} \left(| 0 \rangle +| \psi \rangle  \right) =  \frac{1}{\sqrt{2}} \left(| 0 \rangle -| \psi \rangle  \right),
\end{equation}
if $|\psi \rangle$ is in state $|1\rangle$ in position $k$. Similarly,
\begin{equation}
Z_k \frac{1}{\sqrt{2}} \left(| 0 \rangle -| \psi \rangle  \right) =  \frac{1}{\sqrt{2}} \left(| 0 \rangle +| \psi \rangle  \right).
\end{equation}
If the $Z$ error occurs in the position of where $| \psi \rangle$ is in state $|0 \rangle$ the it has no effect on the GHZ-states. In a fully occupied Fermi-Hubbard lattice model $|\psi\rangle$ has an equal number of qubits in $|0\rangle$ and $|1\rangle$ states, and hence the probability of a single error $Z$ causing the flip between GHZ state is equal to the probability of the state being unaffected. Other types of Pauli noise i.e. $X$ and $Y$ flip will randomize the states $|\psi_0 \rangle$ and $|\psi_\pi \rangle$.

More formally, let $\tilde{W}_0$ be a noisy channel corresponding to noisy circuit $W_0$. The noise is assumed to be Markovian and incoherent. The effect of the noisy channel on the input state $\rho$ is 

\begin{align}
\tilde{W}_0[\rho] \approx & (q+\gamma) W_0 \rho W_0^\dagger + \gamma W_\pi \rho W_\pi^\dagger + (1-q-2\gamma) \frac{I}{2^{2N}}
\end{align}
where $q$ is the probability of an error that randomizes the state. 

The noisy probabilities $p_0^*$ and $p_\pi^*$ are obtained by applying the $\tilde{W}_0$ on  $\rho_0 = |0\rangle \langle 0 |$ and projecting on the state $\Pi_0 = \rho_0 = | 0 \rangle  \langle 0 |$ or $X_j\Pi_0 X_j$. 
\begin{align}
p_0^* = &\text{Tr}[\Pi_0 \tilde{W}_0[\rho_0]  ] \nonumber \\
      = &(q+\gamma) \text{Tr}\left[ \Pi_0 W_0 \rho_0 W_0^\dagger \right]+ \gamma \text{Tr}\left[\Pi_0 W_\pi \rho_0 W_\pi^\dagger \right] \nonumber \\ 
      & +  (1-q-2\gamma) \text{Tr}\left[ \frac{\Pi_0 }{2^{2N}}\right] \nonumber\\
      = & (q+\gamma) p_0 + \gamma p_\pi +\mathcal{K},  \label{eq:p0_star}
\end{align}

where $\mathcal{K}=(1-q-2\gamma)/2^{2N}$

\begin{align}
p_\pi^* = &\text{Tr}[X_j \Pi_0 X_j \tilde{W}_0[\rho_0] ] \nonumber \\
      = &(q+\gamma) \text{Tr}\left[ X_j\rho_0 X_j W_0 \rho_0 W_0^\dagger \right]+ \gamma \text{Tr}\left[X_j \Pi_0 X_j W_\pi \rho_0 W_\pi^\dagger \right] \nonumber \\
      & + (1-q-2\gamma) \text{Tr}\left[ \frac{X_j\Pi_0 X_j }{2^{2N}}\right] \nonumber\\
      = & (q+\gamma) p_\pi + \gamma p_0 +\mathcal{K} \label{eq:ppi_star}
\end{align}    


From, equations (\ref{eq:p0_star}) and (\ref{eq:ppi_star}) it follows that 

\begin{equation}
p_0^* - p_\pi^* =  q (p_0 - p_\pi).    
\end{equation}
Thus, from the measured noisy value of $\textrm{Re}[G(t)]^*=p^*_0 - p^*_\pi$ one can obtain the true value using rescaling $(p^*_0 - p^*_\pi)/q$. 
The factor $q$ can be estimated from the microscopic model of the system. For example, in the presence of two qubit depolarising noise, where gate errors occur with probability $p$, the factor $q$ is given by $q=(1-p)^n$ where $n$ is the number of two-qubit gates in the circuit. This is the approach used to mitigate errors in the experimental results presented in Figure \ref{fig1}.
The scaling parameters $q$ and $\gamma$ can be more accurately obtained by performing Zero-Noisy extrapolation (ZNE) experiment. 

ZNE involves systematically varying the amount of noise in the circuit and from the change in the observed expectation values deducing the scaling parameters. One possible ZNE scheme for measuring $q$ is to use ``folding" i.e. applying \mbox{$\tilde{W}_0^\dagger \circ \tilde{W}_0$}, where $\tilde{W}_0^\dagger$ denotes a noisy reversed circuit $W^\dagger$. The combined map \mbox{$\tilde{W}_0^\dagger \circ \tilde{W}_0$} is equal to identity in the absence of noise. Its effect on the input state \mbox{$\rho_0 = |0 \rangle \langle 0 |$} is

\begin{align}
\tilde{W}^\dagger_0 \circ \tilde{W}_0[\rho_0]  =&
(q+\gamma)^2 W_0^\dagger W_0 \rho_0 W_0^\dagger W_0 + (q+\gamma) \gamma W^\dagger_0 W_\pi \rho_0 W^\dagger_\pi W_0 \nonumber \\
+ & \gamma (q+\gamma) W^\dagger_\pi W_0 \rho_0 W^\dagger_0 W_\pi  + \gamma^2 W_\pi^\dagger W_\pi \rho_0 W_\pi^\dagger W_\pi  \nonumber \\
& +(1-q-2\gamma) \frac{I}{2^{2(N-1)}}. \label{eq:W0W0}
\end{align}
Using $W_0 = V_0^\dagger U V_0$ and $W_\pi = V_0^\dagger U V_\pi$ we have

\begin{equation}
W_0^\dagger W_\pi = V^\dagger_0 U^\dagger V_0 V^\dagger_0 U V_\pi = V^\dagger_0 V_\pi. \label{eq:W0Wpi}
\end{equation}
and
\begin{equation}
W_\pi^\dagger W_0 = V^\dagger_\pi U^\dagger V_0 V^\dagger_0 U V_0 = V^\dagger_\pi V_0. \label{eq:WpiW0}
\end{equation}
Using (\ref{eq:W0Wpi}) and (\ref{eq:WpiW0}), equation (\ref{eq:W0W0}) simplifies to 

\begin{widetext}
\begin{align}
\tilde{W}^\dagger_0 \circ \tilde{W}_0[\rho_0] = &
(q+\gamma)^2  \rho_0  + (q+\gamma) \gamma V^\dagger_0 V_\pi \rho_0 V^\dagger_\pi V_0 
+  \gamma (q+\gamma) V^\dagger_\pi V_0 \rho_0 V^\dagger_0 V_\pi  + \gamma^2 \rho_0 +(1-q-2\gamma) \frac{I}{2^{2(N-1)}} \nonumber \\
 = &  (q+\gamma)^2  \rho_0 + 2\gamma(q+\gamma) X_j \rho_0 X_j + \gamma^2 \rho_0+2(1-q-2\gamma) \frac{I}{2^{2N}}.
\end{align} 
\end{widetext}

After applying \mbox{$\tilde{W}_0^\dagger \circ \tilde{W}_0$}, all qubits are measured in the computational basis. The probability of obtaining the outcome 0 on all qubits i.e. projecting on the state $\rho_0$ is 
\begin{align}
p_0^{\textrm{ZNE}} = &  \textrm{Tr}[\rho_0  \tilde{W}^\dagger_0\left[\tilde{W}_0[\rho_0]\right]] \nonumber \\
= &  (q+\gamma)^2 \textrm{Tr}[\rho_0^2] + 2(q+\gamma)\gamma \textrm{Tr}[\rho_0 X_j \rho_0 X_j] \nonumber \\
& + \gamma^2 \textrm{Tr}[\rho_0^2]+2\mathcal{K}   \nonumber  \\
=  &  (q+\gamma)^2 + \gamma^2+2\mathcal{K}. \label{eq:p0_ZNE} 
\end{align}
Similarly, projecting on state $X_j \rho_0 X_j$ gives 
\begin{align}
p_\pi^{\textrm{ZNE}} = &  \textrm{Tr}[X_j \rho_0 X_j  \tilde{W}^\dagger_0\left[\tilde{W}_0[\rho_0]\right]] \nonumber  \\
= &  (q+\gamma)^2 \textrm{Tr}[X_j \rho_0 X_j\rho_0] + 2\gamma (q+\gamma) \textrm{Tr}[X_j \rho_0 X_j X_j \rho_0 X_j]  \nonumber \\
& + \gamma^2 \textrm{Tr}[X_j \rho_0 X_j\rho_0] +2\mathcal{K} \nonumber  \\
=  &  2\gamma(q+\gamma)+2\mathcal{K}. \label{eq:ppi_ZNE}  
\end{align}
Solving equations (\ref{eq:p0_ZNE}) and (\ref{eq:ppi_ZNE}) for $q$ and $\gamma$ gives

\begin{align}
q= & \sqrt{p_0^{\textrm{ZNE}}-p_\pi^{\textrm{ZNE}}} \label{eq:q_zne} \\
\gamma = & -q + \sqrt{q^2+(p_\pi^{\textrm{ZNE}}-2\mathcal{K})} \label{eq:gamma_zne} 
\end{align}

In practice, ZNE approach involves performing ``folding" experiments to evaluate $q$ and $\gamma$ using equations (\ref{eq:q_zne}) and (\ref{eq:gamma_zne}), which are then used to obtain noiseless $p_0$ and $p_\pi$ from $p_0^*$ and $p_\pi^*$ using equation (\ref{eq:p0_star}) and (\ref{eq:ppi_star}).

\begin{figure}
    \centering
    \includegraphics[scale=0.5]{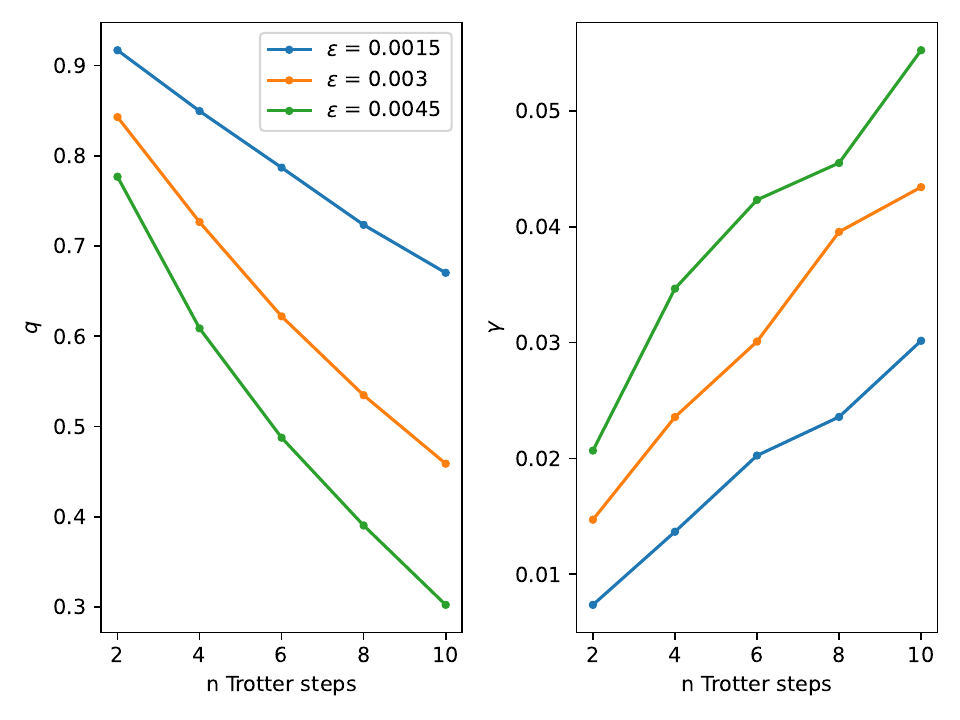} 
    \caption{Measuring $q$ and $\gamma$ factors from numerical simulations of ZNE experiments. The two qubit gate noise is modeled as depolarizing channel with the probability of gate error was set to $\epsilon = 0.0015, 0.003, 0.0045$. 
    The simulation were carried out for $2\times 2$ Fermi-Hubbard model at half-filling with $U=2.0$, $J=0.5$. The final time is set to $T=1.6$ and the number of Trotter step is varied thereby changing the depth of the circuit. Each data point is obtained from 5000 shots.}
    \label{fig:q_gamma_factor}
\end{figure}

\begin{figure}
    \centering
     \includegraphics[scale=0.5]{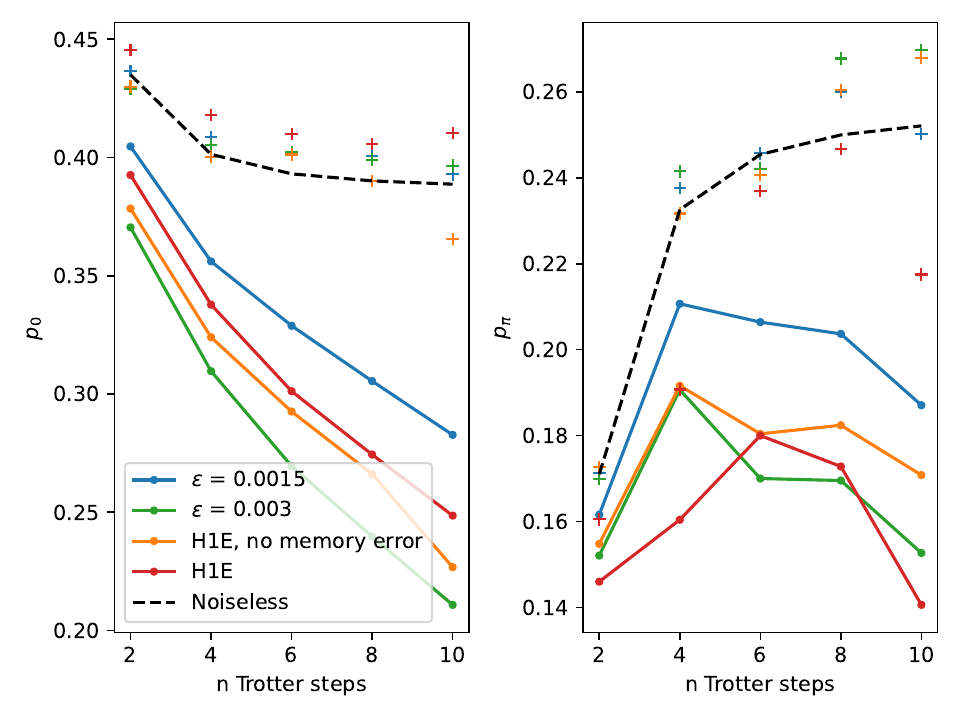} 
    \caption{Numerically testing rescaling error mitigation for $p_0$ and $p_\pi$ on Aer and H1E simulator. Aer simulator was used with depolarising noise model with probability of two qubit error $\epsilon = 0.0015$ and 0.003. H1E emulator was used with the default device-realistic error model and a model with memory error switched off.  
    The parameters $q$ and $\gamma$ were extracted using  simulated ZNE experiments.  The simulation were carried out for $2\times 2$ Fermi-Hubbard model at half-filling with $U=2.0$, $J=0.5$ and the final time $T=1.6$. Each data point is obtained from 5000 shots.}
    \label{fig:rescaling_aer}
\end{figure}

We have tested this procedure numerically by performing noisy circuit simulations on an qiskit Aer backend \cite{Qiskit}, where all two qubit gates experience a uniformly depolarizing noise with probability of gate error $\epsilon$. The simulations were carried out for a $2 \times 2$ Fermi-Hubbard model with 2 spin-up and 2-spin down fermions with $|\psi \rangle = |1010\rangle \otimes |0101\rangle$. The depth of the circuit is varied by changing the number of Trotter steps, while keeping the final time $T$ constant. To apply a rescaling procedure, the factors $q$ and $\gamma$ are obtained using the folding procedure. The resulting $q$ and $\gamma$ factors are shown in Figure \ref{fig:q_gamma_factor}. As expected the $q$ factor decreases with the number of trotter steps (and hence noisy gates), as well as the probability of gate error $\epsilon$. The factor $\gamma$ is much smaller than $q$, since it arises only from the phase errors in the GHZ state preparation circuit. Figure \ref{fig:rescaling_aer} presents probabilities $p_0^*$ and $p_\pi^*$ obtain in noisy simulations and the rescaled values using the measured $q$ and $\gamma$ factors. The rescaled values are appear to be close to the real values for a range of circuit depth, which supports the validity of our model.

The rescaling procedure also works for biased Pauli noise. Figure \ref{fig:rescaling_aer} shows the rescaling error mitigation for simulations on H1E, which models realistic noise in the H1 device. On H1E the Pauli errors affecting the two-qubit gates are not equally likely – in particularly the Z errors are more likely than X and Y errors. In addition, H1E includes the coherent memory errors. Since the coherent errors are not included in the error mitigation model, we expect that the memory errors will degrade the performance of the rescaling procedure. Indeed, from Figure \ref{fig:rescaling_aer}, we can see that the rescaling performs better when the coherent memory error is switched off. As the number of qubits and the depth of the circuit grows, the memory errors will become much more significant and have to be accounted for in the error mitigation procedure. Dynamical decoupling methods can be used to reduced the amount of accumulated memory errors. In addition, the rescaling model can be modified to include extra parameter associated with the build-up of memory error as discussed in the next section.

\section{Memory error}
\label{sec:memory_error}
Let us define:
\begin{equation}
 \rho_{0/\pi}(t) = U(t) V_{0/\pi} \rho_0V_{0/\pi} U(t)^{\dagger}
\end{equation}
The effect of memory error alone can be though of as:
\begin{equation}
\rho_{0/\pi}(t)^* = \int d \theta p(\theta) |\Omega^{0/\pi}_\theta(t) \rangle \langle \Omega^{0/\pi}_\theta(t)| 
\end{equation}
with $\Omega_{\theta}^{0/\pi}(t) = \frac{1}{\sqrt{2}} \left(|\phi \rangle \pm e^{i \theta}e^{iEt} |\psi \rangle \right)$.
This leads to:
\begin{equation}
p_{0/\pi}= \frac{q}{4}(1+|G(t)|^2 \pm ( 2 \langle \cos(\theta)\rangle \Re G(t)-2 \langle \sin(\theta)\rangle \Im G(t)))
\end{equation}
If we call $W_0[\cdot]^*$ the noisy channel which takes into account only memory error and $\tilde{W}_0[\rho]$ the channel which takes into account both depolarizing noise and memory error, we have:
\begin{equation}
\tilde{W}_0[\rho] \approx  (q+\gamma) W_0^* \rho W_0^{*\dagger} + \gamma W_\pi^* \rho W_\pi^{*\dagger} + (1-q-2\gamma) \frac{I}{2^{2N}},
\end{equation}
according to the previous section. 
Therefore we find:
\begin{equation}
\begin{split}
p_{0/\pi}=\left (1-q-2 \gamma \right) \frac{I}{2^N}+\frac{q+ 2 \gamma}{4}\left (1+|G(t)|^2 \right) \\
\pm \frac{q}{4} \left (2 \langle \cos(\theta)\rangle \Re G(t)-2 \langle \sin(\theta)\rangle \Im G(t)\right). 
\end{split}
\end{equation}
Note that in the previous section, we outlined how the effect of depolarizing noise could be mitigated through a simple zero-noise extrapolation procedure. In principle, similar schemes could be developed in the presence of memory error, but since there are three parameters ($\theta$, $q$ and $\gamma$), the procedure would likely be more complicated. This difficulty could potentially be avoided using dynamical-decoupling techniques.
\section{Trotter error analysis}
\begin{figure}
\includegraphics{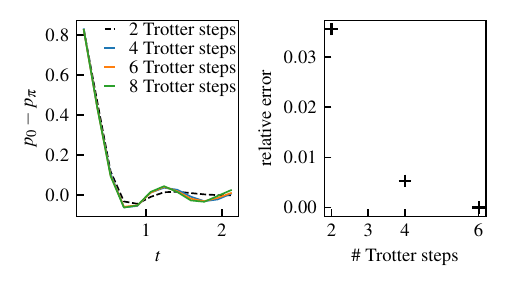}
\caption{\label{fig:trotter} Left: Real part of the Loschmidt amplitude $\Re (G(t)e^{i E t})$ with $E=1.0$ as function of time for different number of Trotter steps. Right: Relative error on the final filtered density of states $D$ as a function of the number of Trotter steps, with the reference value being with 8 trotter steps.}
\end{figure}
The results of the time series as well as the error on the corresponding filtered density of states are presented in Fig. \ref{fig:trotter}. All simulation where performed using the WII method \cite{Zalatel2015}, with $dt=0.025$ and $\chi=300$.

\section{Purification results}

\begin{figure}
    \centering
    \includegraphics{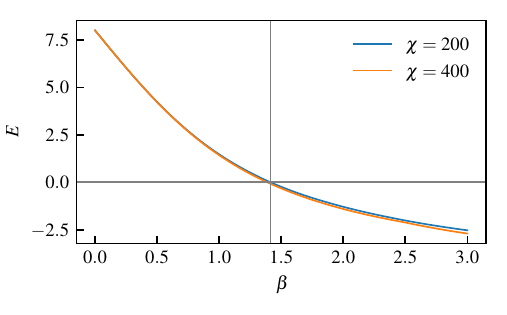}
    \caption{Average value of the energy as a function of the inverse temperature obtained through purification.}
    \label{fig:purification}
\end{figure}

In order compare the minimum energy reachable with product state with other studies performed in the canonical ensemble, we have performed simulation of the system using the purification algorithm \cite{Verstraete2004,Barthel2009} using the TeNPy library \cite{Hauschild2018} and measured the expectation value of the Hamiltonian as function of the temperature. The results are presented on Fig. \ref{fig:purification}.

\section{Convergence of the Markov-chain with bond-dimension}
\label{sec:bond_dimension}
\begin{figure}
\includegraphics{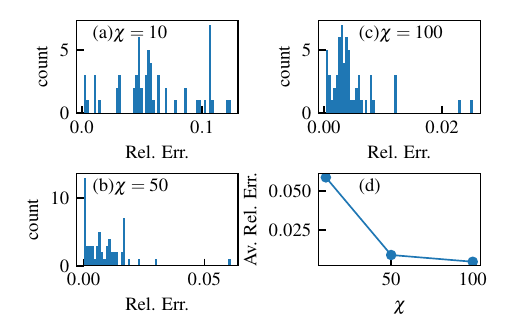}
\caption{Convergence of the filtered density of states for a set of states as a function of the bond dimension. The states are the first 60 samples of a Markov chain obtained with $\chi=200$. Panels (a) to (c): histograms of the relative error on the filtered density of states for $\chi=10, 50$, and $100$, respectively. Panel (d): convergence of the average of the relative errors on the filtered density of states as a function of the bond dimension. The reference value in calculating relative error is the one corresponding to $\chi=200$.}
\label{fig:chi_convergence}
\end{figure}

When checking the bond-dimension convergence of the MPS simulation, we found that low-bond dimensions are sufficient to capture the filtered density of states 
with high accuracy. We show the relative error (compared to $\chi=200$) on the filtered density of states for 60 different states on Fig. \ref{fig:chi_convergence}, at $E=1.0$. For $\chi=100$, which we use in the main text, the average relative error is below $1\%$ with very few outliers around $2\%$.

\section{Influence of the cut-off time and measurement frequency on the filtered density of states}
\label{sec:cutoff}
In order to quantify the error on the truncation of the series of Eq. \eqref{eq:series}, we calculated the value of the filtered density of state with $x=0.5$, for the N\'eel state and $E=1.0$, with 8 Trotter steps and $\alpha=2 \sqrt{L}$. We find $D=0.2468$ with $x=1.0$ and  $D=0.2447$ for $x=0.5$, indicating a good convergence in time. Similarly, we tested the dependence on measurement by doubling $\alpha$. With $\alpha=4 \sqrt{L}$, we find a result almost identical result  $D=0.2469$, indicating a good convergence with the number of measurements.

\section{Second-order scaling for first-order trotterization of Loschmidt amplitudes}\label{app:trotter_scaling}

It is known that the error of first-order trotterization generally scales as $ \mathcal{O}(t^2/n)$, while the error of second-order Trotterizaton scales as $\mathcal{O}(t^3/n^2)$, where $t$ is the total time and $n$ is the number of Trotter steps.
In the following, we prove that for Hamiltonians whose non-commuting terms are all \textit{real} in some basis,  then for all  \textit{real} wavefunctions in the same basis, first-order trotterization of its Loschmidt amplitudes scales rather as $\mathcal{O}(t^3/n^2)$.
Note the notion of realness is basis dependent and this proposition is true as long as there exists some basis in which all quantities are real.

Let the Hamiltonian be $H = A + B$, where $A$ and $B$ are non-commuting operators with commutator $C \coloneqq [A,B]$.
Using the Baker–Campbell–Hausdorff formula, we can write the first-order trotterization as: 
\begin{equation}\label{eq:U}
	U(t, n) \coloneqq [e^{-i A \frac{t}{n}} e^{-i B\frac{t}{n}}]^n = e^{-it(A+B) - \frac{t^2}{n} \frac{C}{2} + \mathcal{O}\left(\frac{t^3}{n^2}\right)} 
\end{equation}
The Taylor series of a function $ f(\delta) = e^{X+\delta Y}$ reads
\begin{equation}
	f(\delta) = e^{X} + \delta\  \int_0^1 d\alpha\ e^{\alpha X} Y e^{(1-\alpha) X} + \mathcal{O}(\delta^2)\;.
\end{equation}
Using this formula to expand the RHS of \eqref{eq:U} as a function of small time steps $\delta \coloneqq \frac{t}{n}$ with $X = -itH$ and $Y = - t C /2 + \mathcal{O}(t^2/n)$, we get
\begin{equation}
	U(t, n) = e^{-it H} - \frac{t^2}{n} \int_0^1 d\alpha e^{- i tH \alpha}  C e^{- i tH (1-\alpha)}+ \mathcal{O}(t^3/n^2)\; .
\end{equation}
The leading error term in the Loschmidt amplitude $\braket{\psi| e^{-it H} |\psi}$ would then be $\frac{t^2}{n} \epsilon$ with
\begin{align}
	\epsilon &\coloneqq \int_0^1 d\alpha \braket{\psi|  e^{- i tH \alpha}  C e^{- i tH (1-\alpha)} |\psi} \\
	&= \int_{-\half}^{\half} d\alpha \braket{\psi|  e^{- i tH (\alpha+\half)}  C e^{- i tH (\half-\alpha)} |\psi}\\
	&= \int_{0}^{\half} d\alpha \braket{\psi|  e^{- i tH (\alpha+\half)} C e^{- i tH (\half-\alpha)} +  e^{- i tH (\half-\alpha)}  C  e^{- i tH (\half+\alpha)} |\psi}\\
& = \int_{0}^{\half} d\alpha \braket{\psi(t_1))| C | \psi(t_2) } + \braket{\psi(-t_2))| C | \psi(-t_1) } \\
&= \int_{0}^{\half} d\alpha \braket{\psi(t_1))| C | \psi(t_2) } - \braket{\psi(-t_1))| C| \psi(-t_2) }^* 
\end{align}
where $t_1\coloneqq - t (\alpha+\half)$ and $t_2 \coloneqq t (\half-\alpha)$, and we used the fact that $C$ is anti-hermitian (because it is the commutator of two hermitian operators).
Let us assume the operators $A$ and $B$ are real in some basis $\ket{a}$, then for any  wavefunction that is real in that basis, we have the time-reversal symmetry $\braket{a| \psi(t)} = \braket{a|\psi(-t)}^*$. Moreover, the commutator's matrix elements $\braket{a^\prime|C|a}$ are also  real is the same basis. 
Therefore
\begin{equation}
	\begin{split}
 \braket{\psi(-t_1))| C| \psi(-t_2) }^* &= \left[\sum_{a, a^\prime}  \braket{\psi(-t_1))|a^\prime} \braket{a^\prime| C|a} \braket{a|\psi(-t_2) }\right]^*\\ &=  \braket{\psi(t_1)| C| \psi(t_2) } 
	\end{split}
\end{equation}
and the leading error term $\epsilon$ vanishes. The next error term in Loschmidt amplitude is then proportional to $\mathcal{O}(t^3/n^2)$.

Extending this proof to Hamiltonians with more than two terms is straightforward by replacing the commutator $C$ with the sum of all commutators.
\end{appendix}


\clearpage

\bibliography{bib}

\small
\newpage

\clearpage
\onecolumngrid

\end{document}